\documentclass[11pt]{article}

\usepackage{amsmath,amsthm,amsfonts,amssymb}
\title{Topology change in (2+1)-dimensional gravity with non-Abelian Higgs
field}

\author{Alexander I. Nesterov\thanks{Departamento de F\'\i sica,
Universidad de Guadalajara, Guadalajara, Jalisco, M\'exico.
E-mail:  nesterov@cencar.udg.mx }}

\begin{document}
\date{~}
\maketitle
\begin{abstract}
    We study topology change in (2+1)D gravity coupling with
non-Abelian $SO(2,1)$ Higgs field from the point of view of
Morse theory. It is shown that the Higgs potential can be identified as
a Morse function. The critical points of the latter ({\em i.e.} loci of
change of the spacetime topology) coincide with zeros of the Higgs field.
In these critical points two-dimensional metric becomes degenerate, but the
curvature remains bounded.
\end{abstract}

PACS number(s): 04.20.Gz, 02.40.-k   \\

\section{Introduction}

     It is known \cite{Ger}  that if one has any (time orientable) Lorentzian metric
which interpolates between two compact spacelike surfaces of different
topology, then there must exist either closed timelike curves or
singularities and from the Tipler's theorem \cite{Tipl} it follows
that the spacetime which interpolates between topologically different
boundaries cannot be compact, in other words, metric must be ill-defined
somewhere.

    {\em What kind of singularities is it necessary to change topology?}
Horowitz \cite{Hor} has shown that if the metric becomes degenerate on a
set of measure zero, but the curvature remains bounded, there exist smooth
solutions of Einstein's equations in which the topology of space changes.

    We explore the topology change in (2+1)D gravity coupling with
non-Abelian $SO(2,1)$ Higgs field from the point of view of
Morse theory. Our starting point is the first-order Palatini action in
terms of a triad $e^a{}_\mu$, Lorentz connection $\omega^a{}_{b\mu}$ and a
real scalar field $\phi^a$ (section 2). In section 3 we consider the ground
state solutions of the field equations. Finally, in section 4 we show that
the Higgs potential can be identified as a Morse function and the critical
points of latter ({\em i.e.} loci of change of the spacetime topology)
coincide with zeros of the Higgs field.

\section{(2+1)-D Palatini action with matter coupling}

In the Palatini formalism the fundamental variables are the base 1-forms
 $\mbox{\boldmath $e$}^a$ and the Lorentz connection
$\mbox{\boldmath $\omega$}^a{}_b$. These variables are assumed to satisfy
the following relations
\[
g^{\mu\nu} e^a{}_\mu e^b{}_\nu=\eta^{ab},
\quad e^a{}_\mu e^b{}_\nu\eta^{ab}= g_{\mu\nu},\quad
\omega_{ab\mu}= - \omega_{ba\mu},
\]
where $\eta^{ab}$ is the constant Minkowski metric and $g_{\mu\nu}$ is a
spacetime metric ($\mu,\nu$ are spacetime indices and $a,b$
internal indices, both running from 0 to 2; we assume signature
being $(-,+,+)$ and define the Ricci tensor as
$R_{\mu\nu}=R^\lambda{}_{\mu\lambda\nu}$).\footnote{For any internal vector
$\rho^a$ on the spacetime the covariant derivative $D_\mu$ is defined by
\quad $D_\mu\rho^a =\rho^a{}_{,\mu} +\omega^a{}_{b\mu}\rho^b.$ \quad On
external vectors it reduces to partial derivative, $D_\mu v^\nu=
v^\nu{}_{,\mu}$, and $D_\mu e^a{}_\nu= e^a{}_{\nu,\mu} +
\omega^a{}_{b\mu}e^b{}_{\nu}=\Gamma^\alpha_{\nu\mu}e^a{}_\alpha$.}

The Einstein--Hilbert action
\[
I_g=\int\sqrt{-g}\; R\; d^3x
\]
in the Palatini  formalism is given by \cite{Hor,Asht,Unru}
\begin{equation}
I_g=\int{\mbox{\boldmath $e$}^a{\wedge R^{bc}}}\epsilon_{abc},
\end{equation}
where
$\mbox{\boldmath$R$}=d\mbox{\boldmath$\omega$}+\mbox{\boldmath$\omega$}\wedge
\mbox{\boldmath$\omega$}$ is the curvature form of the three-dimensional
Lorentz connection and $\epsilon_{abc}$ is  the constant antisymmetric
tensor.

Varying with respect to $e^a{}_\mu$ one gets the field equations
\begin{equation}
\epsilon_{abc}\epsilon^{\mu\nu\rho}R^{bc}{}_{\nu\rho}=0,
\end{equation}
which are exactly the vacuum Einstein equations,
$\sqrt{-g}G^{a\mu}=\sqrt{-g} G^{\nu\mu} e^a_\nu=0$
($G^{\mu\nu}$ being the Einstein tensor).

Variation with respect to $\omega^a{}_b$ leads to the equations
\begin{equation}
D_{[\mu} e^a{}_{\nu]}=0
\end{equation}
which just determine the torsion--free connection and are equivalent to
the first Cartan structure equations
$d {\bf e}^a+\mbox{\boldmath$\omega$}^a{}_b\wedge {\bf e}^b=0.$

We consider gravity coupled to a real matter field $\mbox{\boldmath
$\phi$}$,
\begin{equation}
{\cal L}_m
=-\sqrt{-g}\left(\frac{1}{2}D_{\mu}\mbox{\boldmath $\phi$}{\cdot D^{\mu}
\mbox{\boldmath $\phi$}} + V(\mbox{\boldmath $\phi$})\right),
\end{equation}
where
\[
V(\mbox{\boldmath $\phi$})=\frac{\gamma}{4}\left(\mbox{\boldmath $\phi$}
{\cdot \mbox{\boldmath$\phi$}}+\frac{\mu}{\gamma}\right)^2
\]
is the Higgs potential,
$\mbox{\boldmath $\phi$}{\cdot\mbox{\boldmath $\phi$}}=
\eta_{ab}\phi^a\phi^b$ and $\sqrt{-g}={\rm det}\|e^a_\mu\|$.

Variation of the total action
\[
I=\frac{1}{2\kappa}I_g+I_m
\]
yields the field equations
\begin{align}
&\frac{1}{\sqrt{-g}}D_\mu(\sqrt{-g}D^\mu\phi^a)=
\frac{\delta V}{\delta \phi_a},&\quad
\epsilon^{\mu\nu\lambda}D_{\nu} e^a{}_{\lambda}=
\kappa\sqrt{-g}\epsilon^{abc} \phi_b D^\mu\phi_c,  \nonumber \\
&\epsilon_{abc}\epsilon^{\mu\nu\rho}R^{bc}{}_{\nu\rho}=
4\kappa{\cal J}^\mu_a,&
\label{eq.1}
\end{align}
where the current density ${\cal J}^\mu_a$ is
given by ${\cal J}^\mu_a=\sqrt{-g}T^{\mu\nu}e_{a\nu}$ and $T^{\mu\nu}$ is
the energy-momentum tensor:  \[ T_{\mu\nu}=D_{\mu}\mbox{\boldmath
$\phi$}{\cdot D_{\nu} \mbox{\boldmath $\phi$}}-
g_{\mu\nu}\left(\frac{1}{2}D_{\lambda}\mbox{\boldmath $\phi$}{\cdot
D^{\lambda} \mbox{\boldmath $\phi$}} + V(\mbox{\boldmath $\phi$})\right).
\]

\section {Ground state solutions}

Let us consider solutions of the field equations in the neighborhood of
the spacelike surface $\Sigma_g $ of genus $g$ ($g>1$). The metric of
spacetime can be written as
\[
d s^2 =-N^2 d t^2 + g_{ij}(dx^i+N^i dt) (dx^j+N^j dt),
\]
where $N,N^i$ and $g_{ij}$ are the lapse function, shift vector and spatial
metric induced on the spacelike hypersurface $\Sigma_g$, respectively. It
is convenient to introduce a Gaussian normal coordinate system, which is
specified by the following conditions: $N=1, \; N^i=0$.

It is well known that  one can always choose a conformally flat metric for
the 2-surface,
\[
g_{ij} = \Omega \delta_{ij},
\]
where $\Omega$ is the strictly positive conformal factor. Setting
$\Omega=e^{f+\bar f}$, we find
\begin{equation}
ds^2=-dt^2+2e^{f+\bar f}dzd\bar{z},
\end{equation}
$z=(x+i y)/{\sqrt 2}, \; \bar z=(x-i y)/{\sqrt
2}$ being the complex coordinates and $f=f(t,z,\bar z)$.  For torsion--free
connection field equations (\ref{eq.1}) are written as
\begin{align}
&\frac{1}{2}\Delta(f+\bar f) +
\left(\frac{1}{2}D_0{\mbox{\boldmath $\phi$}}\cdot D_0{\mbox{\boldmath
$\phi$}} -e^{-(f+\bar f)} D_z{\mbox{\boldmath $\phi$}}\cdot D_{\bar
z}{\mbox{\boldmath $\phi$}}+ V(\phi)-{|\dot{f}|^2}\right)e^{f+\bar
f}=0,\label{eq.2.0} \\
&{\dot f}_{,z}-D_0{\mbox{\boldmath $\phi$}}\cdot
D_z{\mbox{\boldmath $\phi$}}=0, \qquad \bigl(e^ f\bigr)^{\cdot\cdot}+
e^f\Bigl(\frac{1}{2}D_0\mbox{\boldmath $\phi$}{\cdot D_0 \mbox{\boldmath
$\phi$}} -V(\mbox{\boldmath $\phi$})\Bigr )=0,\\
&\epsilon_{abc}\phi^b
D_z\phi^c=0,  \qquad D_z{\mbox{\boldmath $\phi$}}\cdot D_z{\mbox{\boldmath
$\phi$}}=0, \\
& e^{-(f+\bar f)}(D_zD_{\bar{z}}{\phi^a}+D_{\bar{z}}D_z{\phi^a})-
\ddot{\phi}^a-\dot{\phi}^a(\dot{f}+\dot{\bar f})-
\phi^a(\mu+\gamma\mbox{\boldmath $\phi$}{\cdot \mbox{\boldmath
$\phi$}})=0,
\label{eq.2}
\end{align}
where $\Delta=2\partial_z\partial_{\bar z}$ is a two-dimensional Laplacian
(a dot denotes partial derivative with respect to $t,\;
{f}_{,z}=\partial_z {f}$, and we set $\kappa=1$).

Further it is convenient to introduce a complex scalar field
$\mbox{\boldmath$\phi$}=(\varphi^0,\varphi,\overline\varphi)$
defined by
\[
\varphi^0=\phi^0, \;
\varphi=(1/{\sqrt 2})(\phi^1+i\phi^2),\; \overline\varphi=(1/{\sqrt
2})(\phi^1-i\phi^2).
\]
The covariant derivative of $\mbox{\boldmath$\phi$}$ is given by
\begin{align}
&D_z\varphi=\varphi_{,z}-iA_z\varphi+A^0_z\varphi^0,\qquad
D_z\bar\varphi=\bar\varphi_{,z}+iA_{z} \bar\varphi, \qquad
D_z\varphi^0=\varphi^0_{,z}+A^0_z\bar\varphi, \nonumber\\
&D_{\bar z}\bar\varphi= \overline{D_z\varphi},\qquad
D_{\bar z}\varphi^0=\overline{D_z\varphi^0}, \qquad
D_{\bar z}\varphi= \overline{D_z\bar\varphi},
\end{align}
where $A_z=i\bar f_{,z},\; A_{\bar z}=\bar A_z, A^0_z=\dot{f}e^{f}$.

For {\em ground state}  defined as $D_{\mu}\phi^a =0$
($\delta V/\delta \phi_a=0$) the system of field equations
(\ref{eq.2.0})-(\ref{eq.2}) reduces to
\begin{align}
&\Delta(f+\bar f) -2 \bigl(V(\phi)-{|\dot f|^2}\bigr)e^{f+\bar f}=0,\\
&{\dot f}_{,z}=0,\;\ddot{f}+\dot{f}^2 - V(\phi)=0, \\
&\phi^a(\mu+\gamma\mbox{\boldmath $\phi$}{\cdot \mbox{\boldmath $\phi$}})=0,
\label{eq.3}\\
&D_{\bar z}\varphi=0, \quad D_z\varphi=0,\quad D_z\varphi^0=0.
\end{align}

Eq.(\ref{eq.3}) yields two cases:
\[
{\rm A.}\quad \mbox{\boldmath$\phi$}{\cdot \mbox{\boldmath$\phi$}}=
-\frac{\mu}{\gamma}\quad
\Longrightarrow \quad V=0, \qquad
{\rm B.}\quad \phi^a=0 \quad \Longrightarrow  \quad
V=V_0=\frac{\mu^2}{4\gamma}.
\]

\subsection*{Case A. $\mbox{\boldmath$\phi$}{\cdot\mbox{\boldmath$\phi$}}=
-\mu/\gamma, \quad V=0$}

The solution of field equations is given by
\begin{align}
&\varphi^0=\sqrt{\frac{\mu}{\gamma}}\left(\frac{1+|F(z)|^2}{1-|F(z)|^2}\right),
\quad \varphi=\sqrt{\frac{\mu}{\gamma}}\left(\frac{2F(z)|F(z)_{,z}|}
{F(z)_{,z}(1-|F(z)|^2)}\right){e^{i(\gamma^n\bar \xi_n+\bar\gamma^n\xi_n)}},
\nonumber  \\
&f=\ln\left({\frac{\sqrt{2}t|F(z)_{,z}|}
{(1-|F(z)|^2)}}\right) + i(\gamma^n\bar\omega_n+\bar\gamma^n\omega_n),
\quad
\end{align}
where $F(z)$ is an arbitrary function satisfying
$|F(z)|< 1$, $\gamma^n$ is a complex $g$-component vector,
\[
\xi_n=\int^z_{z_0}\mbox{\boldmath$\omega$}_n,
\]
and $\mbox{\boldmath$\omega$}_n=\omega_n dz$ is a holomorphic harmonic
1-form with the standard normalization
\[
\int_{\alpha_m}\mbox{\boldmath$\omega$}_n=\delta_{mn},\quad
\int_{\beta_m}\omega_n=\tau_{mn},
\]
$\alpha_m,\beta_m$ being a canonical homology basis (or closed loops
around handles on $\Sigma_g$) and imaginary part of $\tau$ a positive
matrix.

To summarize, the spacetime metric is given by
\begin{equation}
ds^2=-dt^2+t^2\frac{4|dF|^2}{(1-|F(z)|^2)^2}.
\label{eq.4}
\end{equation}
It is well known that (\ref{eq.4}) is a Poincar\'e metric for the
Riemann surface $\Sigma_g$ ($g>1$). Changing $t\rightarrow 2/\tau$ one can
easily reduce this metric to the vacuum solution \cite{Hos}.

\subsection*{Case B. $\phi^a=0, \quad V=V_0=\mu^2/{4\gamma}$}

    In this case one obtains
\begin{equation}
f=\ln\left({\frac{\sqrt{2}\sinh(\lambda t) |F(z)_{,z}|}
{\lambda(1-|F(z)|^2)}}\right),
\end{equation}
where $\lambda=\sqrt{V_0}$. This leads to the following expression for the
spacetime metric
\begin{equation}
ds^2=-dt^2+\left(\frac{\sinh(\lambda t)}
{\lambda}\right)^2\frac{4|dF|^2}{(1-|F(z)|^2)^2}.
\end{equation}
In the limit $\lambda\rightarrow 0$ this metric reduces to (\ref{eq.4}).

    We note that the energy-momentum tensor $T_{\mu\nu}=-g_{\mu\nu}V_0$,
and the obtained solution can be interpreted as a vacuum solution with the
cosmologocal constant $\Lambda=\kappa V_0$.

\section{Topology change. Application of Morse theory}

    To make the description self-consistent it worth outlining some basic
facts of the Morse theory \cite {M,W,F}. A smooth function $F(x)$ on the
manifold $\frak M$ is called Morse function if all critical points of this
function are non degenerate ones. It means that in the critical points
$\partial F/{\partial x^\alpha}=0$, but ${\rm det}\|\partial^2
F/{\partial x^\alpha\partial x^\beta}\|\neq 0$. The Poincar\'e--Hopf
theorem states that the Euler characteristic $\chi (\frak M)$ of $\frak M$
is related to these critical points by
\begin{equation} \chi(\frak M) =
\sum_{p_i}{\rm sign} \left({\rm det} \left\|\frac{\partial^2 F}{\partial
x^\alpha\partial x^\beta}\right\|\right) , \label{Eul}
\end{equation}
where the sum is taken over all critical points $p_i$. Eq. (\ref{Eul}) shows
that this sum is a topological invariant of manifold independent of
function $F$.  The Morse function near the critical point $p_0$ (suppose
that $x^\alpha(p_0)=0$) is given by
\[
F(x)=F_0-x^2_1-\dots -x^2_\lambda+x^2_{\lambda+1}+\dots+x^2_n,
\]
where $\lambda$ is called {\em the index of the critical point}. The
hypersurface $F(x)=c={\rm const}$ is called {\em level surface} $\frak M_c$
of $F(x)$.  If it contains the critical point,  $\frak M_c$ will be called
{\em a critical level surface} of $F_c$.

    Topology change can be described using the Morse function. Namely, if
$\frak N$ is a differentiable manifold, $F$ a Morse function and
$\frak{M,\;M'\subset N}$ level manifolds of $F$ separated by one
non-degenerate critical point, then $\frak M'$ is obtained from $\frak M$
by a spherical modification.

A notaion of a spherical modification can be illustrated by the following
example. Let $\frak M$ be a two-sphere $S^2$ and take zero-dimensional
sphere $S^0\subset S^2$. $S^0$ is a set of two points and has a
neighborhood consisting of two disjoint disks, $U=S^0\times D^2$. Evidently
$S^2- {\rm Int}\;U$ is a sphere with two holes in it, the boundary being
$S^0\times S^1$. On the other hand, $S^0\times S^1$ is also a boundary
of the cylinder $E^2\times S^1$. The union $S^2-{\rm Int}\;U$ and
$E^2\times S^1$, when ends of the cylinder are attached to the
circumferences of the two holes, is a differentiable manifold $\frak M'$
which is a sphere with one handle, or in other words a torus. It is easy to
construct the reverse operation and obtain two-sphere from a torus by a
spherical modification.

    This situation is quite general. For instance, the Riemann surface
$\Sigma_{g+1}$ can be obtained from $\Sigma_g$, where $g$ is the genus of
$\Sigma_g$, by spherical modification corresponding near the critical point
of index 1 to the Morse function
\[
F(x,y,z)=c-z^2+x^2+y^2 .
\]
Here the critical level is determined by $F(x,y,z)=c$ and $\Sigma_g$
by $F(x,y,z)=c-\epsilon$, while $\Sigma_{g+1}$ is determined by
$F(x,y,z)=c+\epsilon$. The inverse process $\Sigma_{g+1}
\longrightarrow\Sigma_{g}$, annihilation of a handle, is described by
the Morse function
\[
F(x,y,z)=c+z^2-x^2-y^2
\]
corresponding to the critical point of index 2.

    Assuming ${\rm det}\|\partial{\phi^a}/{\partial x^\beta}\|\neq 0$,
let us introduce a new coordinate system $\{X^i\}$:
\[
X=\phi^1(x^\alpha),\quad
Y=\phi^2(x^\alpha),\quad
Z=\phi^0(x^\alpha).
\]
Then the equation $V(X^i(x^\alpha))={\rm const}$  determines
embedding $\Sigma_g\longrightarrow E^3$. Let $\phi^a=0$ in the point
$p_0\in\Sigma_g$. Near to $p_0$ the Morse function defined
as $F(X^i)=V(\phi)$ is given by
\[
F= F_0 + \frac{\mu}{2}(-Z^2+X^2+Y^2),
\]
and $p_0$ is the critical point of index 1 if $\mu>0$ (or index 2, if
$\mu<0$).

\section{Concluding remarks}

For the solutions obtained in the section 3 the critical points of the
Higgs potential are in ({\bf A}) degenerates critical points, in ({\bf B})
nondegenerates ones. Thus, only in case {\bf B} it is possible to identify
the Higgs potential as a Morse function. The critical points $p_i$ of the
latter ({\em i.e.} loci of change of the spacetime topology) coincide with
zeros of the Higgs field, $\phi^a(p_i)=0$. From field equations it follows
that near to these points the two-dimensional metric is $g_{z\bar z}\sim
(\phi^1)^2+(\phi^2)^2$, but the curvature remains bounded.

The scenario of topology change can be as follows. Far from the critical
points the spaceime metric takes the form (17) and the topology of
spacetime is hold by the scalar field $\mbox{\boldmath$\phi$}$.  Near the
zeros of the Higgs field the spacetime metric is given by the expression
(19). After topology change the spacetime metric takes the same form (19)
but it is considered as the vacuum solution with the cosmological constant
$\Lambda=\kappa V_0$. Thus the appearence of the cosmological constant is
the result of topology change.

The relation between topology change and Morse theory found here can be
seen as a more general phenomen and expand to four-dimensional spacetime.
This work is being developed.

\section*{Acknowledgments}

This essay was selected for an Honorable Mention by the Gravity Research
Foundation, 1996. The work was supported by CONACyT Grant 1626P-E.


\end{document}